\documentclass[11pt]{article}

\usepackage{amsmath,amssymb}
\usepackage{psfrag}
\usepackage{graphicx}

\makeatletter

 \@addtoreset{equation}{section}
\makeatother

\newcommand{\vev}[1]{\langle {#1} \rangle}

\newcommand{\del}{\partial}

\newcommand{\mbf}{\boldsymbol}

\newcommand{\LS}{\ \ \ \ \ \ \ \ \ \ }
\newcommand{\ls}{\ \ \ \ \ }
\newcommand{\wt}{\widetilde}

\newcommand{\ol}{\overline}
\newcommand{\ul}{\underline}

\newcommand{\kahler}{K\"{a}hler }

\newcommand{\bsubeq}{\begin{subequations}}
\newcommand{\esubeq}{\end{subequations}}

\newcommand{\noi}{\noindent}

\renewcommand{\d}{{\rm d}}

\newcommand{\nn}{\nonumber}

\newcommand{\C}{{\mathbb C}}
\newcommand{\Z}{{\mathbb Z}}

\newcommand{\CO}{{\cal O}}

\renewcommand{\P}[1]{\C{\bf P}^{#1}}

\newcommand{\e}{{\rm e}}
\newcommand{\all}{{}^{\forall}}
\newcommand{\exist}{{}^{\exists}}
\newcommand{\F}{{\mathbb F}}
\newcommand{\B}{{\mathbb B}}
\newcommand{\R}{{\mathbb R}}

\newcounter{Enumerate}

\begin{document}

\allowdisplaybreaks{

\setcounter{page}{0}

\begin{titlepage}

{\normalsize
\begin{flushright}
KEK-TH-886, 
OU-HET 442\\
hep-th/0305072\\
May 2003
\end{flushright}
}

\vspace{15mm}

\begin{center}
{\Huge \kahler Potentials on Toric Varieties}

\vspace{11mm}

\bigskip
{\Large Tetsuji {\sc Kimura}}

\vspace{5mm}

{\sl
Theory Division, Institute of Particle and Nuclear Studies,\\
KEK, High Energy Accelerator Research Organization\\
Tsukuba, Ibaraki 305-0801, Japan}

and 

{\sl 
Department of Physics,
Graduate School of Science, Osaka University, \\
Toyonaka, Osaka 560-0043, Japan}

\vspace{2mm}

{\tt tetsuji@post.kek.jp}

\end{center}

\vspace{15mm}


\begin{abstract}
One has believed that low energy effective theories of 
the Higgs branch of gauged linear sigma
models correspond to supersymmetric nonlinear sigma models, which have
been already investigated by many works.
In this paper we discuss a explicit derivation of supersymmetric
nonlinear sigma models from gauged linear sigma models.
In this process we construct \kahler potentials of some
two-dimensional toric varieties explicitly.
Thus we will be able to study some algebraic varieties in the
language of differential geometry. 
\end{abstract}

\end{titlepage}


\newpage

\section{Introduction}

Investigations of string theory on curved backgrounds become more and
more important tasks to understand the dynamics and dualities of string
theory and M-theory. 
One of the most important strategies is 
studying and understanding (supersymmetric) nonlinear sigma models
with conformal symmetries whose target spaces are non-trivially
curved.
But it is so difficult to discuss a direct investigation of 
nonlinear sigma models that we consider alternative models which become
nonlinear sigma models at some specific limits.
One of the powerful model is a gauged linear sigma model (GLSM).

GLSM is defined as one class of ${\cal N}=2$
supersymmetric gauge theory in two-dimensional spacetime.
This is one of the most important tools to
investigate and understand the string theory dynamics and topological
aspects of string theory.
GLSM was introduced by Witten in order to analyze supersymmetric field
theories whose target spaces are represented by the local descriptions
of toric varieties \cite{W93}.

Even though the formulation of GLSM is very simple,  
this gauge theory has various important properties.
First, in the Higgs branch of GLSM, we believe that 
the low energy effective theory becomes ${\cal N}=2$ supersymmetric
nonlinear sigma model (NLSM) whose target space is a toric variety.
In the Coulomb branch 
we can trace an effect of worldsheet instantons \cite{OV02}.
Moreover we can investigate some phases of GLSM by changing
Fayet-Iliopoulos (FI) parameters.
If we set FI parameters as positive under one specific condition
(i.e., the Calabi-Yau condition or the chiral anomaly free condition), 
the low energy effective theory becomes the conformal sigma model on
Calabi-Yau manifold.
If we set them as negative, Landau-Ginzbrug model is realized in the
low energy limit.
Due to this phenomenon, we can discuss the Calabi-Yau/Landau-Ginzbrug
correspondence easily.

GLSMs are applied to a lot of topics in field theory and string
theory: topological field theory and topological string \cite{W93, MP94}, 
conifold transitions \cite{SGM}, closed string
tachyon condensations \cite{V01}, 
mirror symmetry \cite{HV00} and mirror symmetry 
with D-branes \cite{HIV00, GJS}, and more.

${\cal N}=2$ supersymmetric NLSMs in two-dimensional spacetime and
their gauge theory extensions have been 
investigated by many works for a long time
because these models have various important phenomena such as
asymptotic freedom, confinements, mass gaps in the spectra, and so on.
First, supersymmetric NLSM on projective space 
$\P{N}$ was studied in order to understand 
chiral symmetry breaking and quark confinements \cite{DiV}.
The non-abelian extension to Grassmannian was considered by
\cite{Aoyama}.
In recent years these models were generalized to 
supersymmetric NLSMs on hermitian symmetric spaces by \cite{HN}. 

It is important to understand the dynamics of 
${\cal N}=2$ supersymmetric NLSMs on (non)compact 
Calabi-Yau manifolds in order to 
understand the dynamics of superstring theories on curved backgrounds.
Thus we constructed \kahler potentials and metrics of 
$\CO(-N)$ on $\P{N-1}$ as noncompact Calabi-Yau $N$-folds \cite{Ca},
and canonical line bundles on Einstein-K\"{a}hler manifolds \cite{HKN} 
in the language of supersymmetric NLSMs.
These potentials will be applied to classical and quantum
supergravity and superstring theories on these curved backgrounds 
and we will understand some new properties of dynamics on curved
spacetime.
Moreover, we should discuss dynamics of supergravity and string
theories on other curved spacetime in order to understand string
dualities.

In this paper,
we will construct
\kahler potentials of some two-dimensional toric varieties, which are
not Calabi-Yau manifolds.
First we will review the projective space $\P{2}$, which is the basic
varieties of other two-dimensional toric varieties.
Second, we will construct Hirzebruch surfaces $\F_k$ ($k=0,1,2$), which
are $\P{1}$ bundles on weighted projective space $W\P{1,1,k}$.
Third, del Pezzo surfaces $\B_k$ ($k=0,1,2$), which is $\P{2}$ blown
up at $k$ points, will be introduced.
Since these toric varieties are powerful tools to investigate string
duality,
it is of important for us to understand their \kahler potentials and
some objects derived from \kahler potentials in the language of
differential geometry as well as algebraic geometry.
If we understand some aspects of toric varieties in the language of 
differential geometry,
we will be able to investigate the dynamics of supergravity and
string/M-theory beyond the discussions of topological or
non-dynamical aspects of these theories.

The organization of this paper is as follows:
In section two we review gauged linear sigma models and
supersymmetric NLSMs. And we consider a relation between these
models. 
In section three we construct \kahler potentials of Hirzebruch surfaces
and del Pezzo surfaces in detail.
We give some comments on metrics and other geometrical objects.
In section four we discuss some applications of these \kahler
potentials.
In appendix \ref{fan} we discuss an introduction of two-dimensional
toric varieties.


\section{Gauged Linear Sigma Model}

Let us construct GLSM \cite{W93}.
This is one class of ${\cal N}=2$ supersymmetric gauge theories in
two-dimensional spacetime.
GLSM consists of chiral superfields $\Phi^i$, anti-chiral superfields
$\ol{\Phi}{}^i$, and vector superfields $V_a$ with complexified
$U(1)_a$ symmetry.
Lagrangian is written as 
\begin{align}
{\cal L}_{\rm GLSM} 
\ &= \ 
\int \! \d^4 \theta \, 
\Big\{ \sum_a - \frac{1}{e_a^2} \ol{\Sigma}{}_a \Sigma_a 
+ \sum_{i} \ol{\Phi}{}^i \, \e^{\sum_a 2 Q_i^a V_a} \Phi^i \Big\}
+ \Big( \frac{1}{2\sqrt{2}} \int \! \d \theta{}^+ \d \ol{\theta}{}^-
\, \sum_a \tau_a \Sigma_a  + (h.c.) \Big) 
\; ,  
\label{GLSM}
\end{align}
where $e_a$ are $U(1)_a$ gauge coupling constants, which are
dimensionful parameters.
$Q_i^a$ are
$U(1)_a$ charges of chiral superfields.
We denote a kinetic term of a vector superfield in terms of a
twisted chiral superfield, whose definition is 
$\Sigma_a = \frac{1}{\sqrt{2}} \ol{D}{}_+ D_- V_a$.
$\tau_a$ are complexified FI parameters: $\tau_a = r_a -
i \theta_a/2\pi$, 
where $r_a$ are the FI parameters and $\theta_a$ are $U(1)_a$ theta angles. 

Witten introduced GLSM as a field theory realization of a toric variety.
In this context, chiral superfields are regarded as homogeneous
coordinates of toric variety,
complexified $U(1)$ gauge symmetries as symmetries of algebraic torus
$\C^*$.
And $U(1)$ charges of chiral superfields correspond to Mori vectors of
toric variety.

We would like to make a low energy effective theory of GLSM.
In order to find this,
we should take some vacua of GLSM to be supersymmetric, 
and consider the Higgs branch.
Now let us discuss supersymmetric vacua.
First, we investigate the potential of GLSM.
Under the Wess-Zumino gauge the potential is written as
\begin{align*}
U \ &= \ 
\sum_a \Big\{ 
\frac{e_a^2}{2} 
\Big( r_a - \sum_i Q_i^a |\phi^i|^2 \Big)^2 \Big\}  
+ 2 \sum_{a,b}  \ol{\sigma}{}_a \sigma_b
 \Big( \sum_i Q_i^a Q_i^b \, |\phi^i|^2 \Big) 
\; ,
\end{align*}
where $\phi^i$ and $\sigma$ are the scalar components of the chiral
superfields and vector superfields, respectively. 
$D_a$ are auxiliary fields of $V_a$ whose equations of motion are given as
\begin{align*}
\frac{1}{e_a^2} D_a \ &= \ r_a - \sum_i Q_i^a |\phi^i|^2 \; .
\end{align*}
In order to obtain the supersymmetric vacua, we set the potential and
auxiliary fields as
\begin{align}
U \ &= \ 0 \; , \ls D_a \ = \ 0 \; . \label{constraint}
\end{align}
This condition makes the moduli space of GLSM (\ref{GLSM}).
Next
let us consider the Higgs branch of this model.
In general, one believe that the low energy effective theory of
GLSM become a (supersymmetric) nonlinear sigma model (NLSM) whose target
space is a toric variety. 
In the Higgs branch, 
some chiral superfields have vacuum expectation values.
Thus gauge multiplets become massive by the Higgs mechanism. 
Theta angles $\theta_a$ vanish because the energy of
the supersymmetric ground state must be zero.
In the IR limit ($e_a \to \infty$), 
kinetic terms of gauge multiplets $\frac{1}{e_a^2} \ol{\Sigma}{}_a
\Sigma_a$ decouple from other terms.
So the gauge multiplets do not propagate and freeze the dynamics:
the gauge multiplets in (\ref{GLSM}) become {\sl auxiliary fields}.

In the context of the above discussions,
we start from a ``frozen'' Lagrangian
\begin{align}
{\cal L}_{\rm F} 
\ &= \ 
\int \! \d^4 \theta \, \Big\{
\sum_{i} \ol{\Phi}{}^i \, \e^{\sum_a 2 Q_i^a V_a} \Phi^i
+ \sum_a \big( - 2 r_a \, V_a \big) \Big\} 
\ \equiv \ \int \! \d^4 \theta \, K 
\; , \label{frozen-L}
\end{align}
which does not contain the kinetic terms of vector superfields.
Note that $K$ is a \kahler potential.
Though we integrate out the auxiliary vector superfields from
(\ref{frozen-L}), 
the imaginary part of the complexified gauge symmetry is not fixed.
So we choose one gauge to fix this residual gauge symmetry.
In this process we obtain the supersymmetric NLSM \cite{DiV, Aoyama, HN}.
By choosing one 
gauge we can normalize some components of
homogeneous coordinates.
This process corresponds to the constraint on the homogeneous
coordinates (\ref{constraint}) 
and then we obtain the supersymmetric NLSM in terms of local
coordinates of toric variety.


\section{\kahler Potentials}

In this section 
we construct \kahler potentials of two-dimensional toric varieties.
First, we review the construction of two-dimensional projective space 
in subsection \ref{P2}.
In subsection \ref{F} we apply the technique to construction of
\kahler potentials of Hirzebruch surfaces.
Moreover in subsection \ref{B} 
we construct \kahler potentials of del Pezzo surfaces in the same way.
 
Subsection \ref{F} and \ref{B} are the main parts of this paper.


\subsection{Projective Space} \label{P2}

Let us first review a construction of \kahler potential of
two-dimensional projective space $\P{2}$.
The toric data of $\P{2}$ are described in appendix \ref{fan-def-P2}.

GLSM for $\P{2}$ consists of chiral superfields $\Phi^i$ ($i=1,2,3$),
anti-chiral superfields $\ol{\Phi}{}^i$ and a vector superfield $V$
with the complexified $U(1)$ gauge symmetry.
We assign $U(1)$ charges of chiral superfields as in Table
\ref{charge-P2}:
\begin{table}[h]
\begin{center}
\begin{tabular}{c|ccc}
& $\Phi^1$ & $\Phi^2$ & $\Phi^3$ \\ \hline 
$U(1)$ & 1 & 1 & 1 
\end{tabular}
\caption{\sl $U(1)$ charges.}
\label{charge-P2}
\end{center}
\end{table}

\noi
In the IR limit of GLSM for $\P{2}$, the \kahler potential of the
frozen Lagrangian (\ref{frozen-L}) is written as
\begin{align}
K \ &= \ 
\big( |\Phi^1|^2 + |\Phi^2|^2 + |\Phi^3|^2 \big) \, \e^{2 V} - 2 r V
\; , \label{kahler-P2}
\end{align}
where $V$ is now an auxiliary field.
In order to obtain the \kahler potential of supersymmetric NLSM,
we solve the equation of motion of this auxiliary field:
\begin{align*}
\frac{\del {\cal L}_{\rm F}}{\del V} \ = \ 0 \ \ &: \ls
r \ = \ \big( |\Phi^1|^2 + |\Phi^2|^2 + |\Phi^3|^2 \big) \; , \\
\therefore \ \ \ 
\e^{2 V} \ &= \ \frac{r}{|\Phi^1|^2 + |\Phi^2|^2 + |\Phi^3|^2} \; .
\end{align*}
Note that we can divide $r$ by $\sum_i |\Phi^i|^2$ 
since all chiral superfields do not vanish simultaneously 
(due to the $D$-flatness condition of GLSM). 
Substituting this solution into \kahler potential (\ref{kahler-P2}),
we obtain the potential of supersymmetric NLSM:
\begin{align}
K \ &= \ r \log \big( |\Phi^1|^2 + |\Phi^2|^2 + |\Phi^3|^2 \big) \; .
\end{align}
Note that we neglect a constant term which does not contribute to
the Lagrangian.
Now we should fix a residual gauge symmetry, i.e., an imaginary part of
complex $U(1)$ gauge symmetry.
This process corresponds to dividing other
chiral superfields ($\Phi^1$, $\Phi^3$) by $\Phi^2$ and we normalize
$\Phi^2$ to a constant as follows:
\begin{align}
K (X^a, \ol{X}{}^a) 
\ &= \ r \log \big( 1 + |X^1|^2 + |X^2|^2 \big) \; ,
\label{P2-kahler2}
\end{align}
where $X^1 = \Phi^1/\Phi^2$ and $X^2 = \Phi^3/\Phi^2$ and we neglect
a ``constant'' term $r \log |\Phi^2|^2$ as well as other constant terms.
Note that this dividing process corresponds to choosing a local patch
$U_{\sigma_1}$ of $\P{2}$ (see appendix \ref{fan-def-P2}).
Then we obtain the already known formulation of the \kahler potential
of $\P{2}$.
It is easy to construct the \kahler metric of this space.
The metric is defined as $g_{a\ol{b}} = \del_a \ol{\del}{}_b K (x,
\ol{x})$, where $x^a$ are scalar components of chiral superfields
$X^a$. 
This metric is the Fubini-Study metric:
\begin{align}
g_{a \ol{b}} \ &= \ \frac{r}{(1 + |x^1|^2 + |x^2|^2)^2}
\Big\{ (\delta_{a}^1 \, \delta_{b}^1 
+ \delta_{a}^2 \, \delta_{b}^2 ) \big( 1 + |x^1|^2 + |x^2|^2 \big) 
- \big( \delta_a^1 \, \ol{x}{}^1 + \delta_{a}^2 \, \ol{x}{}^2
\big) \big( \delta_b^1 \, x^1 + \delta_{b}^2 \, x^2 \big) \Big\}
\;. 
\end{align}


\subsection{Hirzebruch Surfaces} \label{F}

Let us construct GLSMs for the
Hirzebruch surfaces $\F_k$ ($k = 0, 1, \cdots$) 
by using the toric data (see appendix \ref{fan-def-F}).
In terms of GLSMs, 
scalar components of chiral superfields
represent homogeneous coordinates of toric varieties.
Thus we sometimes call these fields ``homogeneous coordinates.''
In addition,
two abelian vector superfields $V_1$ and $V_2$ 
are introduced in order to make the algebraic torus $(\C^*){}^2$. 
We denote complexified gauge groups as $U(1)_1^{\C} \times U(1)_2^{\C}$.
Chiral superfields $\Phi^i$ ($i=1,2,\cdots,4$) 
have charges of these gauge groups as in Table \ref{charge-F}:
\begin{table}[h]
\begin{center}
\begin{tabular}{c|cccc}
 & $\Phi^1$ & $\Phi^2$ & $\Phi^3$ & $\Phi^4$ \\ \hline 
$U(1)_1$ & 1 & 1 & $k$ & 0 \\
$U(1)_2$ & 0 & 0 & 1 & 1 
\end{tabular}
\caption{\sl $U(1) \times U(1)$ charges.}
\label{charge-F}
\end{center}
\end{table}

\noi
Substituting this data into (\ref{frozen-L}) 
we obtain the frozen GLSM for this variety, whose \kahler potential is
written as
\begin{align}
K \ &= \ 
|\Phi^1|^2 \, \e^{2 V_1} + |\Phi^2|^2 \, \e^{2 V_2} 
+ |\Phi^3|^2 \, \e^{2 k V_1 + 2 V_2} + |\Phi^4|^2 \, \e^{2 V_2}  \nn \\
\ & \ \ \ \ 
- 2 r_1 V_1 - 2 r_2 V_2 \; , \label{kahler-F}
\end{align}
where $r_i$ are FI parameters and we assume $r_1 > r_2$ for the
positive definiteness of the volume of target space manifold.


\vspace{5mm}

\noi
\ul{$k=0$ case: $\F_0 = \P{1} \times \P{1}$}

The simplest Hirzebruch surface is $\F_0$, which is globally equal to
the direct product of projective spaces: $\F_0 = \P{1} \times
\P{1}$. 
The \kahler potential of this surface is well known and we can obtain
from (\ref{frozen-L}) by integrating out the vector superfields as 
\begin{align} 
K (X^a, \ol{X}{}^a) 
\ &= \ 
r_1 \log \big( 1 + |X^1|^2 \big) + r_2 \log \big( 1 + |X^2|^2 \big) 
\; , \label{P2-kahler}  
\end{align}
where chiral superfields 
$X^1 = \Phi^1/\Phi^2$ and $X^2 = \Phi^3/\Phi^4$ correspond to
the local coordinates of $U_{\sigma_1}$ in $\F_0 = \P{1} \times \P{1}$. 
Note that we identify the local patch $U_{\sigma_i}$ to the patches on
toric cone $\sigma_i$ in Figure \ref{fan-Hir}.
Since 
the local coordinates of other patches and transition rules among
them are described in appendix \ref{fan-def-F},
it is sufficient for us 
to consider the \kahler potential and metric only in one local
patch, for example, as (\ref{P2-kahler}) in $U_{\sigma_1}$.
We obtain the metric of $\F_0$ by
differentiating the \kahler potential in terms of the local
coordinates $x^a$ ($a = 1, 2$):
\begin{align}
g_{a \ol{b}} 
\ &= \ 
\frac{\del}{\del x^a} \frac{\del}{\del \ol{x}{}^b} K (x, \ol{x})  
\ = \ 
\delta_{a}^1 \, \delta_b^1 \frac{r_1}{(1 + |x^1|^2)^2} 
+ \delta_a^2 \, \delta_b^2 \frac{r_2}{(1 + |x^2|^2)^2} 
\; . \label{P2-metric}
\end{align}
Note that $x^a$ are scalar components of chiral superfields $X^a$, which
denote the local coordinates of the target space of NLSM.
This metric (\ref{P2-metric}) 
corresponds to the sum of the Fubini-Study metrics, which has
been already known.



\vspace{5mm}

\noi
\ul{$k=1$ case: $\F_1$}

Here we construct the metric of $\F_1$, 
which is the simplest variety 
that one $\P{1}$ is non-trivially fibered by the other $\P{1}$.

The equations of motion for the auxiliary vector superfields are
obtained from (\ref{frozen-L}) as:
\bsubeq \label{eq1}
\begin{align}
\frac{\del {\cal L}_{\rm F}}{\del V_1} \ = \ 0 \ \ &: \ls
r_1 \ = \ \big( |\Phi^1|^2 + |\Phi^2|^2 \big) \, \e^{2 V_1} + |\Phi^3|^2 \,
\e^{2 V_1 + 2 V_2} \; , \\
\frac{\del {\cal L}_{\rm F}}{\del V_2} \ = \ 0 \ \ &: \ls
r_2 \ = \ |\Phi^3|^2 \, \e^{2 V_1 + 2 V_2} + |\Phi^4|^2 \, \e^{2 V_2}
\; . 
\end{align}
\esubeq
The solutions are obtained as
\bsubeq \label{sol1}
\begin{align}
\e^{2 V_1} \ &= \ 
\frac{1}{2 A} \Big\{ - B + \sqrt{B^2 - 4 A C} \, \Big\} 
\; , \\
\e^{2 V_2} \ &= \ 
\frac{2 r_2 (|\Phi^1|^2 + |\Phi^2|^2)}
{2 |\Phi^3|^2 (|\Phi^1|^2 + |\Phi^2|^2) - B + \sqrt{B^2 - 4 A C}} 
\; ,   
\end{align}
\esubeq
where
\begin{align*}
A \ &= \ \big( |\Phi^1|^2 + |\Phi^2|^2 \big) |\Phi^3|^2 \; , \\
B \ &= \ \big( |\Phi^1|^2 + |\Phi^2|^2 \big) |\Phi^4|^2 - (r_1 - r_2)
|\Phi^3|^2 \; , \\
C \ &= \ - r_1 |\Phi^4|^2 \; . 
\end{align*}
Note that these solutions include all region of the Hirzebruch surface
$\F_1$, i.e., some zero points $\Phi^3 = 0$ and
$\Phi^4=0$ are also included. 
These zero points are the origin and the infinity point of the
$\P{1}$ fiber.

Substituting these solutions (\ref{sol1}) into the \kahler potential
(\ref{kahler-F}), we obtain the \kahler potential of the NLSM of $\F_1$:
\begin{align}
K 
\ &= \
- r_1 \log \Big\{ \frac{1}{2 A} \big( - B + \sqrt{B^2 - 4 AC} \big)
\Big\} \nn \\
\ & \ \ \ \ 
- r_2 \log \Big\{
\frac{2 t (|\Phi^1|^2 + |\Phi^2|^2)}{2 (|\Phi^1|^2 + |\Phi^2|^2)
  |\Phi^4|^2 - B + \sqrt{B^2 - 4 AC}} \Big\} \nn \\
\ & \ \ \ \ 
- r_2 \, \frac{-B + \sqrt{B^2 - 4 AC}}{2 (|\Phi^1|^2 + |\Phi^2|^2)
  |\Phi^4|^2 - B + \sqrt{B^2 - 4 AC}} 
\; . \label{F1-kahler}
\end{align}
Note that the third term comes from the non-trivial fibration of
$\P{1}$.
Chiral superfields $\Phi^i$ are related to each other because we
should fix the residual gauge symmetry (imaginary part of $U(1)^{\C}
\times U(1)^{\C}$). 

As we will discuss in appendix \ref{fan-def-F},
we represent Hirzebruch surfaces using four local patches $U_{\sigma_i}$
($i = 1, 2, \cdots, 4$).
Thus we will define local coordinates on these patches and consider
relationships among them where some patches overlap.

Now let us define the local coordinates of various patches in terms of
homogeneous coordinates $\Phi^i$. 
In the same way as $\F_0$, we describe local coordinates as follows:
\begin{align*}
(X^1, X^2) \ &= \ \Big( \frac{\Phi^1}{\Phi^2} , \frac{\Phi^3}{\Phi^2 \Phi^4}
  \Big) \ \in \ U_{\sigma_1} \; , \ls 
\mbox{where} \ \ \Phi^2,  \Phi^4 \ \neq \ 0 
\; .
\end{align*}
Since transition rules of coordinates among local patches are
described in appendix \ref{fan-def-F},
we denote the \kahler
potential and metric of $\F_1$ in the local patch $U_{\sigma_1}$ only.

Let us write down the \kahler potential of $\F_1$ 
in terms of the local coordinates $X^a$ as
\begin{align}
K (X^a, \ol{X}{}^a) \ &= \ 
r_1 \log \wt{D}{}_+
+ r_2 \log \Big\{ 1 + 
\frac{\wt{D}{}_-}{2 (1 + |X^1|^2)} \Big\} 
- r_2 \, \frac{\wt{D}{}_-}{2 (1 + |X^1|^2) + \wt{D}{}_-} 
\; . \label{F1-kahler2}
\end{align}
Note that we neglect constant terms because they do not
appear in the Lagrangian. 
In (\ref{F1-kahler2}) we use the following function:
\begin{align*}
\wt{B} \ &= \ (1 + |X^1|^2) - (r_1 - r_2) |X^2|^2 \; , \ls 
\wt{D}{}_{\pm} \ = \ \pm \wt{B} + \sqrt{\wt{B}{}^2 + 4 (1 + |X^1|^2)
  |X^2|^2} \; .
\end{align*}
Thus we have obtained the Lagrangian of
the supersymmetric NLSM on the Hirzebruch surface $\F_1$.

Let us give some comments on the geometrical objects.
We obtain the \kahler metric as
\begin{align}
g_{a \ol{b}} \ &= \ \frac{\del}{\del x^a} \frac{\del}{\del \ol{x}{}^b}
K (x, \ol{x}) \nn \\
\ &= \ 
- \delta_a^1 \, \delta_b^1 \frac{r_2}{(1 + |x^1|^2)^2}
+ \frac{r_1}{\wt{D}{}_+} \del_a \ol{\del}{}_b \wt{D}{}_+
- \frac{r_1}{(\wt{D}{}_+)^2} \del_a \wt{D}{}_+ \, \ol{\del}{}_b
\wt{D}{}_+ \nn \\
\ & \ \ \ \ 
+ \delta_a^1 \, \delta_b^1 \, \frac{2 r_2}{2 (1 + |x^1|^2) +
  \wt{D}{}_-} \nn \\
\ & \ \ \ \ 
+ \frac{r_2}{[2 (1 + |x^1|^2) + \wt{D}{}_-]^2}
\Big\{ 2 \delta_a^1 \, \delta_b^1 \wt{D}{}_- 
+ \wt{D}{}_- \, \del_a \ol{\del}{}_b \wt{D}{}_- 
+ \del_a \wt{D}{}_- \, \ol{\del}{}_b \wt{D}{}_- 
- 4 \delta_a^1 \, \delta_b^1 \, |x^1|^2 \Big\} \nn \\
\ & \ \ \ \ 
- \frac{2 r_2 \wt{D}{}_-}{[2 (1 + |x^1|^2) + \wt{D}{}_-]^3}
\big\{ 2 \delta_a^1 \, \ol{x}{}^1 + \del_a \wt{D}{}_- \big\}
\big\{ 2 \delta_b^1 \, x^1 + \ol{\del}{}_b \wt{D}{}_- \big\} \; ,
\end{align}
where 
\begin{align*}
\del_a \wt{B} &= \delta_a^1 \, \ol{x}{}^1 - (r_1 - r_2)
\delta_a^2 \, \ol{x}{}^2 \; , \ls
\ol{\del}{}_b \wt{B} = \delta_b^1 \, x^1 - (r_1 - r_2)
\delta_b^2 \, x^2 \; , \\
\del_a \ol{\del}{}_b \wt{B} &= 
\delta_a^1 \, \delta_b^1 - (r_1 - r_2) \delta_a^2 \,
\delta_b^2 \;  , \\
\del_a \wt{D}{}_{\pm} &=
\pm \del_a \wt{B} 
+ \frac{\wt{B} \del_a \wt{B} + 2 r_1 [ \delta_a^1 \, \ol{x}{}^1
  |x^2|^2 + \delta_a^2 \, \ol{x}{}^2 (1 + |x^1|^2)]}
{\sqrt{\wt{B}{}^2 + 4 r_1 (1 + |x^1|^2) |x^2|^2}} \; , \\
\ol{\del}{}_b \wt{D}{}_{\pm} &= 
\pm \del_a \wt{B} 
+ \frac{\wt{B} \ol{\del}{}_b \wt{B} + 2 r_1 [ \delta_b^1 \, x^1
  |x^2|^2 + \delta_b^2 \, x^2 (1 + |x^1|^2)]}
{\sqrt{\wt{B}{}^2 + 4 r_1 (1 + |x^1|^2) |x^2|^2}} \; , \\
\del_a \ol{\del}{}_b \wt{D}{}_{\pm} &=  
\pm \del_a \ol{\del}{}_b \wt{B} \\ 
\ & \ \ \ 
+ \big\{ \wt{B}{}^2 + 4 r_1 (1 + |x^1|^2) |x^2|^2 \big\}^{-1/2} \\ 
\ & \ls \times 
\Big\{ \wt{B} \del_a \ol{\del}{}_b \wt{B} + \del_a \wt{B} \ol{\del}{}_b \wt{B} 
+ 2 r_1 \big[ \delta_a^1 \delta_b^1 |x^2|^2 + \delta_a^2
  \delta_b^2 (1 + |x^1|^2) + \delta_a^1 \delta_b^2 \ol{x}{}^1 x^2
+ \delta_a^2 \delta_b^1 x^1 \ol{x}{}^2 \big] \Big\} \\ 
\ & \ \ \ 
- \big\{\wt{B}{}^2 + 4 r_1 (1 + |x^1|^2) |x^2|^2 \big\}^{-3/2} \\ 
\ & \ls \times
\Big\{ \wt{B} \del_a \wt{B} + 2 r_1 [ \delta_a{}^1 \, \ol{x}{}^1
  |x^2|^2 + \delta_a^2 \, \ol{x}{}^2 (1 + |x^1|^2)] \Big\} \\ 
\ & \ls \times 
\Big\{ \wt{B} \ol{\del}{}_b \wt{B} + 2 r_1 [ \delta_b{}^1 \, {x}{}^1
  |x^2|^2 + \delta_b^2 \, {x}{}^2 (1 + |x^1|^2)] \Big\} \; . 
\end{align*}
In principle, we can explicitly obtain curvature tensors, Ricci tensors and the
Euler number of the Hirzebruch surface $\F_1$ 
from the \kahler potential (\ref{F1-kahler2}).


\vspace{5mm}

\noi
\ul{$k=2$ case: $\F_2$}

Now we consider another Hirzebruch surface, i.e., $\F_2$ ($k=2$).
In terms of the toric data written in Table \ref{charge-F},
the \kahler potential of the frozen Lagrangian (\ref{frozen-L}) is written as
\begin{align}
K
\ &= \ 
|\Phi^1|^2 \, \e^{2 V_1} + |\Phi^2|^2 \, \e^{2 V_1} 
+ |\Phi^3|^2 \, \e^{4 V_1 + 2 V_2} + |\Phi^4|^2 \, \e^{2 V_2} 
- 2 r_1 \, V_1 - 2 r_2 \, V_2 \; . \label{F2}
\end{align}
Then the equations of motion for $V_1$ and $V_2$ are obtained as
\bsubeq \label{eq2}
\begin{align}
\frac{\del {\cal L}_{\rm F}}{\del V_1} \ = \ 0 \ \ &: \ls
r_1 \ = \ \big( |\Phi^1|^2 + |\Phi^2|^2 \big) \, \e^{2 V_1} + 2 |\Phi^3|^2 \,
\e^{4 V_1 + 2 V_2} \; , \\
\frac{\del {\cal L}_{\rm F}}{\del V_2} \ = \ 0 \ \ &: \ls
r_2 \ = \ |\Phi^3|^2 \, \e^{4 V_1 + 2 V_2} + |\Phi^4|^2 \, \e^{2 V_1}
\; .
\end{align}
\esubeq
The solutions are very complicated:
\bsubeq \label{sol2}
\begin{align}
\e^{2 V_1} \ &= \ 
\frac{E^{2/3} - 4 (3 AC - B^2) - 2 B E^{1/3}}{6 A E^{1/3}}
\; , \\
\e^{2 V_2} \ &= \ 
\frac{r_2 (6 A E^{1/3})^2}
{(6 AE^{1/3})^2 |\Phi^4|^2 + |\Phi^3|^2 [ E^{2/3} - 4 (3 AC - B^2) - 2
    B E^{1/3}]^2} 
\; ,
\end{align}
\esubeq
where 
\begin{align*}
A \ &= \ 
(|\Phi^1|^2 + |\Phi^2|^2) |\Phi^3|^2 \; , \ls
B \ = \ - (r_1 - 2 r_2) |\Phi^3|^2 \; , \\
C \ &= \ (|\Phi^1|^2 + |\Phi^2|^2) |\Phi^4|^2 \; , \ls
D \ = \ - r_1 |\Phi^4|^2 \; , \\
E \ &= \ 36 ABC - 108 A^2 D - 8 B^3 \\ 
\ & \ \ \ \ 
+ 12 \sqrt{3} A \big( 4 AC^3 - B^2 C^2 - 18 ABCD + 27 A^2 D^2 + 4 B^3
D \big)^{1/2} \; .  
\end{align*}
The \kahler potential is obtained as
\begin{align}
K 
\ &= \
- r_1 \log \Big\{ \frac{E^{2/3} - 4 (3 AC - B^2) - 2 B E^{1/3}}{6 A
  E^{1/3}} \Big\} \nn \\
\ & \ \ \ \ 
- r_2 \log \Big\{ \frac{r_2 (6 A E^{1/3})^2}
{(6 AE^{1/3})^2 |\Phi^4|^2 + |\Phi^3|^2 [ E^{2/3} - 4 (3 AC - B^2) - 2
    B E^{1/3}]^2} \Big\} \nn \\
\ & \ \ \ \ 
- 2 r_2 \, \frac{|\Phi^3|^2 [E^{2/3} - 4 (3 AC - B^2) - 2 B E^{1/3}]^2}
{(6 AE^{1/3})^2 |\Phi^4|^2 
+ |\Phi^3|^2 [ E^{2/3} - 4 (3 AC - B^2) - 2 B E^{1/3}]^2}
\; . \label{F2-kahler1}
\end{align}
Note that we neglect the constant terms which do not contribute to Lagrangian.
Since the homogeneous coordinates $\Phi^i$ are related to each other,
we choose some local patches to describe the local coordinates of
this variety.
Thus let us define the local coordinates of various patches in terms of
homogeneous coordinates $\Phi^i$. 
In the same way as the previous discussions, 
we describe local coordinates as follows:
\begin{align*}
(X^1, X^2) \ &= \ \Big( \frac{\Phi^1}{\Phi^2} , 
\frac{\Phi^3}{(\Phi^2)^2 \Phi^4} \Big) \ \in \ U_{\sigma_1} \; , \ls 
\mbox{where} \ \ \Phi^2,  \Phi^4 \ \neq \ 0
\; .
\end{align*}
In terms of these local coordinates, 
we can describe the \kahler potential in $U_{\sigma_1}$ as follows:
\begin{align}
K (X^a, \ol{X}{}^a) 
\ &= \ 
r_1 \log \big( 1 + |X^1|^2 \big) + r_1 \log |X^2|^2
+ \frac{r_1}{3} \log \wt{E} 
- r_1 \log \big\{ \wt{E}{}^{2/3} - 4 (3 \wt{A} \wt{C} - \wt{B}{}^2)
- 2 \wt{B} \wt{E}{}^{1/3} \big\} \nn \\
\ & \ \ \ \
+ r_2 \log \Big\{ 1 + \frac{\wt{E}{}^{2/3} - 4 (3 \wt{A}
    \wt{C} - \wt{B}{}^2) - 2 \wt{B} \wt{E}{}^{1/3}}
{36 (1 + |X^1|^2)^2 |X^2|^2 \wt{E}{}^{2/3}} \Big\} \nn \\
\ & \ \ \ \ 
- 2 r_2 \, \frac{\wt{E}{}^{2/3} - 4 ( 3 \wt{A} \wt{C} - \wt{B}{}^2)
- 2 \wt{B} \wt{E}{}^{1/3}}{36 (1 + |X^1|^2)^2 |X^2|^2 \wt{E}{}^{2/3}
+ [\wt{E}{}^{2/3} - 4 (3 \wt{A} \wt{C} - \wt{B}{}^2) - 2 \wt{B}
  \wt{E}{}^{1/3}]^2} \nn \\
\ & \ \ \ \ 
+ \mbox{(constant terms)}
\; , \label{F2-kahler2}
\end{align}
where 
\begin{align*}
\wt{A} \ &= \ (1 + |X^2|^2)|X^2|^2 \; , \ls 
\wt{B} \ = \ - (r_1 - 2 r_2) |X^2|^2 \; , \ls
\wt{C} \ = \ 1 + |X^1|^2 \; , \\
\wt{E} \ &= \ 36 \wt{A}\wt{B}\wt{C} 
+ 108 r_1 \wt{A}{}^2 - 8 \wt{B}{}^3 \\ 
\ & \ \ \ \ 
+ 12 \sqrt{3} \wt{A} 
\big( 4 \wt{A} \wt{C}{}^3 - \wt{B}{}^2 \wt{C}{}^2 
+ 18 r_1 \wt{A} \wt{B} \wt{C} + 27 (r_1)^2 \wt{A}{}^2 - 4 r_1
\wt{B}{}^3 \big)^{1/2} \; .  
\end{align*}
The \kahler potential (\ref{F2-kahler2}) 
seems to give a singular metric at $x^2 =0$ because
of the existence of $r_1 \log |X^2|^2$.
But if one combine this term with others in order to absorb some
divergence at $x^2 =0$
then one find the metric is smooth in all region.


\subsection{del Pezzo Surfaces} \label{B}

In this subsection we consider the \kahler potential of the
del Pezzo surface $\B_k$ which corresponds to $\P{2}$ blown up at $k$
points. 
For simplicity we consider only $k = 0,1,2$ case.
It is very difficult to obtain $\B_3$ and we cannot construct \kahler
potentials $\B_{k \geq 4}$ in the way we discussed in the previous sections. 

Since the simplest del Pezzo surface $\B_0$ is not blown up by $\P{1}$, 
this is the two-dimensional projective space $\P{2}$ itself, 
which we have already discussed in subsection \ref{P2}.
It is well known that the del Pezzo surface $\B_1$ corresponds to 
the Hirzebruch surface $\F_1$ because the toric fan of the former
corresponds to the one of the latter.
So we discuss only $\B_2$ in this subsection.

{}From the toric fan of the del Pezzo surface $\B_2$ 
(see appendix \ref{fan-def-B2}), 
we construct GLSM for $\B_2$ in terms of chiral superfields
$\Phi^i$ ($i = 1, 2, \cdots, 5$) 
and complexified $U(1)^3$ vector superfields $V_a$ ($a = 1, 2, 3$).
The $U(1)^3$ charges are assigned as in Table \ref{charge-B}: 
\begin{table}[h]
\begin{center}
\begin{tabular}{c|ccccc}
 & $\Phi^1$ & $\Phi^2$ & $\Phi^3$ & $\Phi^4$ & $\Phi^5$ \\ \hline 
$U(1)_1$ & 1 & 1 & 1 & 0 & 0 \\
$U(1)_2$ & 0 & 0 & 1 & 1 & 0 \\
$U(1)_3$ & 1 & 0 & 0 & 0 & 1 
\end{tabular}
\caption{\sl $U(1)^3$ charges.}
\label{charge-B}
\end{center}
\end{table}

Substituting the toric data of Table \ref{charge-B} into (\ref{frozen-L}), 
we give the \kahler potential of the supersymmetric Lagrangian:
\begin{align}
K
\ &= \ 
|\Phi^1|^2 \, \e^{2 V_1 + 2 V_3} + |\Phi^2|^2 \, \e^{2 V_1}
+ |\Phi^3|^2 \, \e^{2 V_1 + 2 V_2} + |\Phi^4|^2 \, \e^{2 V_2}
+ |\Phi^5|^2 \, \e^{2 V_3} \nn \\
\ & \ \ \ \ 
- 2 r_1 \, V_1 - 2 r_2 \, V_2 - 2 r_3 \, V_3 
\; . \label{kahler7}
\end{align}
Note that we assume $r_1 \geq r_2$ and $r_1 \geq r_3$ in order for the positive
definiteness of the metric.
We obtain the equations of motion for these vector superfields as
\begin{align*}
\frac{\del {\cal L}_{\rm F}}{\del V_1} \ = \ 0 \ \ &: \ls 
r_1 \ = \ |\Phi^1|^2 \, \e^{2 V_1 + 2 V_3} + |\Phi^2|^2 \, \e^{2 V_1} 
+ |\Phi^3|^2 \, \e^{2 V_1 + 2 V_2}
\; , \\
\frac{\del {\cal L}_{\rm F}}{\del V_2} \ = \ 0 \ \ &: \ls 
r_2 \ = \ |\Phi^3|^2 \, \e^{2 V_1 + 2 V_2} + |\Phi^4|^2 \, \e^{2 V_2} 
\; , \\
\frac{\del {\cal L}_{\rm F}}{\del V_3} \ = \ 0 \ \ &: \ls 
r_3 \ = \ |\Phi^1|^2 \, \e^{2 V_1 + 2 V_3} + |\Phi^5|^2 \, \e^{2 V_3} 
\; .
\end{align*}
The solutions are as follows:
\bsubeq \label{sol3}
\begin{align}
\e^{2 V_1} \ &= \ 
\frac{E^{2/3} - 4 (3 AC - B^2) - 2 BE^{1/3}}{6 A E^{1/3}} \; , \\
\e^{2 V_2} \ &= \ 
\frac{6 r_2 A E^{1/3}}
{6 A E^{1/3} |\Phi^4|^2 + |\Phi^3|^2 [ E^{2/3} - 4 (3 AC - B^2) -
    2 B E^{1/3}]} \; , \\
\e^{2 V_3} \ &= \ 
\frac{6 r_3 A E^{1/3}}
{6 A E^{1/3} |\Phi^5|^2 + |\Phi^1|^2 [ E^{2/3} - 4 (3 AC - B^2) -
    2 B E^{1/3}]} \; ,
\end{align}
\esubeq
where
\begin{align*}
A \ &= \ |\Phi^1|^2 |\Phi^2|^2 |\Phi^3|^2 \;, \\
B \ &= \ |\Phi^2|^2 \big( |\Phi^1|^2 |\Phi^4|^2 + |\Phi^3|^2 |\Phi^5|^2 \big)
+ (-r_1 + r_2 + r_3) |\Phi^1|^2 |\Phi^3|^2 \; , \\
C \ &= \ |\Phi^2|^2 |\Phi^4|^2 |\Phi^5|^2 + (-r_1 + r_2) |\Phi^3|^2 |\Phi^5|^2
+ (-r_1 + r_3) |\Phi^1|^2 |\Phi^4|^2 \; , \\
D \ &= \ - r_1 |\Phi^4|^2 |\Phi^5|^2 \; , \\
E \ &= \  36 ABC - 108 A^2 D - 8 B^3 \\ 
\ & \ \ \ \ 
+ 12 \sqrt{3} A \big( 4 AC^3 - B^2 C^2 - 18 ABCD + 27 A^2 D^2 + 4 B^3
D \big)^{1/2} \; .  
\end{align*}
Substituting (\ref{sol3}) into the \kahler potential (\ref{kahler7}), 
we obtain 
\begin{align}
K 
\ &= \ 
- r_1 \log \Big\{
\frac{E^{2/3} - 4 (3 AC - B^2) - 2 B E^{1/3}}{6 A E^{1/3}} \Big\} \nn \\
\ & \ \ \ \ 
- r_2 \log \Big\{
\frac{6 r_2 A E^{1/3}}{6 A E^{1/3} |\Phi^4|^2 + |\Phi^3|^2 
[E^{2/3} - 4 (3 AC - B^2) - 2 BE^{1/3}]} \Big\} \nn \\
\ & \ \ \ \ 
- r_3 \log \Big\{
\frac{6 r_3 A E^{1/3}}{6 A E^{1/3} |\Phi^5|^2 + |\Phi^1|^2 
[E^{2/3} - 4 (3 AC - B^2) - 2 BE^{1/3}]} \Big\} \nn \\
\ & \ \ \ \ 
- r_2 \, \frac{|\Phi^3|^2 [ E^{2/3} - 4 (3 AC - B^2) - 2 B
    E^{2/3}]}
{6 A E^{1/3} |\Phi^4|^2 + |\Phi^3|^2 [ E^{2/3} - 4 (3 AC - B^2) -
    2 B E^{2/3}]} \nn \\
\ & \ \ \ \ 
- r_3 \, \frac{|\Phi^1|^2 [ E^{2/3} - 4 (3 AC - B^2) - 2 B
    E^{2/3}]}
{6 A E^{1/3} |\Phi^5|^2 + |\Phi^1|^2 [ E^{2/3} - 4 (3 AC - B^2) -
    2 B E^{2/3}]}
\; . \label{B2-kahler1}
\end{align}
Note that
chiral superfields are related to each other under the gauge-fixing.
Now let us discuss the local coordinate systems and the relations
among them.
We define the local coordinates of various patches in terms of
homogeneous coordinates $\Phi^i$. 
We describe local coordinates as follows:
\begin{align*}
(X^1, X^2) \ &= \ 
\Big( \frac{\Phi^1}{\Phi^2 \Phi^5} , \frac{\Phi^3}{\Phi^2 \Phi^4}
  \Big) \ \in \ U_{\sigma_1} \; , \ls 
\mbox{where} \ \ \Phi^2,  \Phi^4, \Phi^5 \ \neq \ 0
\; .
\end{align*}
As we will discussed as Figure \ref{transition-B} 
in appendix \ref{fan-def-B2}, 
we describe transition rules of coordinates among local patches.
Thus it is sufficient for us to 
describe the \kahler potential of $\B_2$ in $U_{\sigma_1}$ as
\begin{align}
K (X^a, \ol{X}{}^a) 
\ &= \ 
(r_1 - r_2 - r_3) \log \wt{A} + \frac{1}{3} (r_1 - r_2 - r_3) \log \wt{E}
\nn \\
\ & \ \ \ \
- r_1 \log \Big\{ \wt{E}{}^2 - 4 ( 3 \wt{A} \wt{C} - \wt{B}{}^2) - 2
\wt{B} \wt{E}{}^{1/3} \Big\} \nn \\
\ & \ \ \ \ 
+ r_2 \log \Big\{ 6 \wt{A} \wt{E}{}^{1/3} 
+ |X^2|^2  \big[ \wt{E}{}^2 - 4 (3 \wt{A} \wt{C} - \wt{B}{}^2) - 2 \wt{B}
  \wt{E}{}^{1/3} \big] \Big\} \nn \\
\ & \ \ \ \ 
+ r_3 \log \Big\{ 6 \wt{A} \wt{E}{}^{1/3} 
+ |X^1|^2  \big[ \wt{E}{}^2 - 4 (3 \wt{A} \wt{C} - \wt{B}{}^2) - 2 \wt{B}
  \wt{E}{}^{1/3} \big] \Big\} \nn \\
\ & \ \ \ \ 
- r_2 \, \frac{|X^2|^2 [ \wt{E}{}^{2/3} - 4 (3 \wt{A} \wt{C} -
    \wt{B}{}^2) - 2 \wt{B} \wt{E}{}^{2/3}]}
{6 \wt{A} \wt{E}{}^{1/3} 
+ |X^2|^2 [ \wt{E}{}^{2/3} - 4 (3 \wt{A} \wt{C} - \wt{B}{}^2) 
- 2 \wt{B} \wt{E}{}^{2/3}]} \nn \\
\ & \ \ \ \ 
- r_3 \, \frac{|X^1|^2 [ \wt{E}{}^{2/3} - 4 (3 \wt{A} \wt{C} -
    \wt{B}{}^2) - 2 \wt{B} \wt{E}{}^{2/3}]}
{6 \wt{A} \wt{E}{}^{1/3} + |X^1|^2 [ \wt{E}{}^{2/3} - 4 (3 \wt{A}
    \wt{C} - \wt{B}{}^2) - 2 \wt{B} \wt{E}{}^{2/3}]} \nn \\
\ & \ \ \ \ 
+ \mbox{(constant terms)}
\; , 
\label{B2-kahler2}
\end{align}
where 
\begin{align*}
\wt{A} \ &= \ |X^1|^2 |X^2|^2 \; , \\
\wt{B} \ &= \ (-r_1 + r_2 + r_3) + |X^1|^2 + |X^2|^2 \; , \\
\wt{C} \ &= \ 1 + (-r_1 + r_3) |X^1|^2 + (-r_1 + r_2) |X^2|^2 \; , \\
\wt{E} \ &= \  36 \wt{A}\wt{B}\wt{C} 
+ 108 r_1 \wt{A}{}^2 - 8 \wt{B}{}^3 \\ 
\ & \ \ \ \ 
+ 12 \sqrt{3} \wt{A} 
\big( 4 \wt{A} \wt{C}{}^3 - \wt{B}{}^2 \wt{C}{}^2 
+ 18 r_1 \wt{A} \wt{B} \wt{C} + 27 (r_1)^2 \wt{A}{}^2 - 4 r_1
\wt{B}{}^3 \big)^{1/2} \; .  
\end{align*}
Thus we have obtained the Lagrangian of the supersymmetric NLSM on $\B_2$.

Let us give one comment on the metric from the \kahler potential 
(\ref{B2-kahler2}).
The metric from (\ref{B2-kahler2}) also seems
to be singular at some points,
but one can arrange (\ref{B2-kahler2}) in order that delta functions
(for example, $\delta^2 (x^2, \ol{x}{}^2)$) do not appear in the metric.
Thus we have constructed a well-defined \kahler potential of $\B_2$.


\section{Discussion}

In this paper we have explicitly constructed \kahler potentials of some
two-dimensional toric varieties, i.e., Hirzebruch surfaces $\F_k$ ($k =
1, 2$) and del Pezzo surface $\B_2$.
In principle we can construct the \kahler metrics, curvature tensors
and more objects on these surfaces from the \kahler potentials.

In the language of algebraic geometry,
one have discussed only topological aspects of string theory on
nontrivial backgrounds, i.e., 
one have not been able to investigate some dynamics of theories.
Let us give some examples.
String theory on toric varieties have been considered in the context
of string duality. 
Iqbal, Neitzke and Vafa investigated a correspondence
between toroidal compactifications of M-theory and del Pezzo surfaces
\cite{INV01}. 
They called this correspondence {\sl a mysterious duality}.
But they considered them only in terms of cohomologies and
exceptional curves in del Pezzo.
There is another example:
String theory with D-branes on curved backgrounds are discussed by 
Hori, Iqbal and Vafa in terms of GLSMs and its mirror dual models
\cite{HIV00}. They also investigated string theory only in terms of
algebro-geometric properties.
By using the \kahler potentials of del Pezzo which we have constructed,
we will directly check this correspondence at supergravity level 
and will be able to obtain deeper insights of the mysterious duality,
and string theory dynamics with D-branes wrapped on non-trivial cycles.

We are able to give one comment on the geometric engineering.
In \cite{HKN}, we have constructed 
\kahler potentials and metrics of the canonical line bundles on
Einstein-K\"{a}hler manifolds as noncompact Calabi-Yau manifolds.
Thus we will be able to construct \kahler potentials on
canonical bundle over toric varieties with vanishing first Chern classes.

\section*{Acknowledgements}

The author would like to thank 
Hiroyuki Fuji,
Tomoyuki Fujita,
Kiyoshi Higashijima, 
Toshio Nakatsu,
Kazuhiro Sakai,
Dan Tomino,
Hiroshi Umetsu,
Yukinori Yasui and 
Takashi Yokono
for valuable comments.
The author would also like to thank all members of the KEK theory group
for a good opportunity to study a lot of things. 
This work is supported in part by JSPS Research Fellowships for Young
Scientists. 


\begin{appendix}

\section*{Appendix}


\section{Toric Fans} \label{fan}

In this appendix we consider complex two-dimensional toric varieties,
which are the projective space $\P{2}$ (a basic
example of toric varieties), 
Hirzebruch surfaces $\F_k$ and del Pezzo surfaces $\B_k$.
We introduce toric fans of these varieties, 
and discuss local patches and their transition rules. 
The arguments presented in this appendix 
are given by the famous TASI lecture by Greene \cite{G96} in detail.

Now let us introduce some definitions.
Let $N = \Z^d$ be an $d$-dimensional lattice and $N_{\R} = N \otimes_{\Z} \R$
an extension of it.
The dual lattice of $N$ and its $\R$ extension are defined by
$M = \Z^d = {\rm Hom} (N, \Z)$ and $M_{\R} = M \otimes_{\Z} \R$, where
${\rm Hom}$ means homomorphism.
We denote the inner product between two vectors ${\mbf n} \in N_{\R}$
and ${\mbf m} \in M_{\R}$ as $\vev{{\mbf m}, {\mbf n}}$.

We define a $d_i$-dimensional 
{\sl strongly convex rational polyhedral cone} $\sigma_i$ 
in $N_{\R}$ as follows:
\begin{align}
{\sigma}_i \cap N_{\R}\ &= \ 
{\R}_{\geq 0} \, {\mbf n}_{i1} + {\R}_{\geq 0} \, {\mbf n}_{i2} + \cdots +
{\R}_{\geq 0} \, {\mbf n}_{i d_i} \; , \label{cones}
\end{align}
where ${\mbf n}_{ij}$ ($j= 1,2,\cdots,d_i$) are elements of $N$ such
that
\begin{align*}
\exist {\mbf m} \ \in \ M_{\R} \; , \ls
\vev{{\mbf m} , {\mbf n}_{ij}} \ > \ 0 \ \ \ \mbox{for} \ \ \all
j \; .
\end{align*}
We refer to strongly convex rational polyhedral cone as simply {\sl cone}.

A fan $\Delta$ is defined as a collection of cones $\sigma_i$ which
satisfy the requirement that the face of any cone in $\Delta$ is also
in $\Delta$.

The dual cones $\check{\sigma}_i$ are defined as
\begin{align}
\check{\sigma}_i \ &= \ 
\big\{ {\mbf m}_i \in M \; ; \ \ \vev{{\mbf m}_i , {\mbf n}_{jk}} \geq 0 \ \
\mbox{for} \ \all {\mbf n}_{jk} \in \sigma_i \big\} 
\; . \label{dual-cones}
\end{align}
Now we choose a finite set of elements ${\mbf m}_{ij} \in M$ such that
\begin{align}
\check{\sigma}_i \cap M_{\R}\ &= \ 
{\R}_{\geq 0} \, {\mbf m}_{i1} + {\R}_{\geq 0} \, {\mbf m}_{i2} + \cdots +
{\R}_{\geq 0} \, {\mbf m}_{i d_i} \; . \label{dual-cones2}
\end{align}
We then find a finite set of some relations as
\begin{align*}
\sum_{j=1}^{d_i} p_{s,j} \, {\mbf m}_{ij} \ &= \ 0 \; , \ls 
p_{s,j} \in \ \Z \; , \ls s \ = \ 1, 2, \cdots, R \; , \\ 
\sum_{j=1}^{d_i} p_j \, {\mbf m}_{ij} \ &= \ 0 \; , \ls
p_j \ = \ \sum_{s=1}^R \mu_s \, p_{s,j} \; , \ls \mu_s \ \in \ \Z \; ,
\end{align*}
where $R$ and $\mu_s$ are some finite integers.
We associate a local coordinate patch $U_{\sigma_i}$ to the cone
$\sigma_i$ by
\begin{align*}
U_{\sigma_i} \ &= \ 
\Big\{ (u_{i1} , u_{i2}, \cdots, u_{i d_i}) \in \C^{d_i} \; ; 
\ \ \ 
\prod_{j=1}^{d_i} (u_{ij})^{p_{s,j}} = 1 \ \ \ \ \mbox{for} \ \ \all s
\Big\} \; , 
\end{align*}
where $u_{ij}$ are local coordinates in $U_{\sigma_i}$.

Now we consider the coordinate transformations where some local
patches overlap such as $U_{\sigma_i} \cap U_{\sigma_j}$.
We prepare a complete set of relations of the form
\bsubeq \label{transit} 
\begin{align}
\sum_{\ell=1}^{d_i} q_{\ell} \, {\mbf m}_{i \ell} 
+ \sum_{\ell=1}^{d_j} q'_{\ell'} \, {\mbf m}_{j \ell'} \ &= \ 0 \; ,  
\ls q_{\ell} , \, q'_{\ell'} \ \in \ \Z \; .
\end{align}
For each of these relations we impose the coordinate transition relations
as
\begin{align}
\prod_{\ell=1}^{d_i} (u_{i \ell})^{q_{\ell}} 
\cdot \prod_{\ell'=1} (u_{j \ell'})^{q'_{\ell'}} \ &= \ 1 \; . 
\end{align}
\esubeq
Using these relations we give some local descriptions of toric varieties.

The $k$-dimensional toric variety 
is also viewed as the \kahler quotient space as
\begin{align}
\frac{\C^d - F_{\Delta}}{(\C^*)^{d-k}} \; ,
\end{align}
where $d$ is the number of one-dimensional cones in the fan $\Delta$,
and $F_{\Delta}$ a subspace of $\C^d$ determined by $\Delta$.
We parametrize $\C^d$ in terms of $d$ homogeneous coordinates $z_i$, 
which are associated with one-dimensional cones $v_i$ in $\Delta$.
Here we introduce the {\sl primitive} generators $n(v_i)$ of $v_i$,
i.e.,
one-dimensional cones are described as $v_i = {\R}_{\geq 0} n (v_i)$.

Now let us define the action of $(\C^*)^{d-k}$ as follows:
\bsubeq \label{torus}
\begin{align}
(\C^*)^{d-k} \ : \ \ z_{i} \ \mapsto \ \prod_{a=1}^{d-n}
  (\lambda_a)^{q_i^a} \, z_{i} \; , \ls
(\lambda_1 , \lambda_2 , \cdots, \lambda_{d-k}) \ \in \ (\C^*)^{d-k}
  \; , 
\end{align}
where $q_i^a$ are integers which satisfy the following linear
relations among $\{n(v_i)\}$:
\begin{align}
\sum_{v_i \in \Delta} q_i^a \, n (v_i) \ &= \ 0 \ls \mbox{for} \ \ 
a \ = \ 1, 2, \cdots, d-k \; .
\end{align}
\esubeq
Local coordinates $\{ u_{ij} \}$ in the local patch $U_{\sigma_i}$ are
represented by homogeneous coordinates as
\begin{align}
u_{ij} \ &= \ 
\prod_{v_{\ell} \in \Delta} z_{\ell}^{\vev{{\mbf m}_{ij}, n(v_\ell)}}
\; . \label{local-homo}
\end{align}
Note that local coordinates are invariant under the $(\C^*)^{d-k}$
action given by (\ref{torus}).


\subsection{Projective space} \label{fan-def-P2}

Let us consider the toric fan which represents the two-dimensional
projective space $\P{2}$. This toric fan consists of various
dimensional cones. Zero-dimensional cone is the origin.
We define one-dimensional cones $v_i$ and two-dimensional cones
$\sigma_i$ as 
\begin{align*}
\left\{
\begin{array}{l}
v_1 \cap N_{\R} \ = \ \R_{\geq 0} \, \vec{n}_1 \\
v_2 \cap N_{\R} \ = \ \R_{\geq 0} (- \vec{n}_1 - \vec{n}_2) \\
v_3 \cap N_{\R} \ = \ \R_{\geq 0} \, \vec{n}_2 
\end{array} \right. 
\; , \ls & \ls 
\left\{
\begin{array}{l}
\sigma_1 \cap N_{\R} 
\ = \ \R_{\geq 0} \, \vec{n}_1 + \R_{\geq 0} \, \vec{n}_2 \\
\sigma_2 \cap N_{\R} 
\ = \ \R_{\geq 0} \, \vec{n}_2 + \R_{\geq 0} (- \vec{n}_1 -\vec{n}_2) \\
\sigma_3 \cap N_{\R} 
\ = \ \R_{\geq 0} \, \vec{n}_1 + \R_{\geq 0} (- \vec{n}_1 -\vec{n}_2) 
\end{array} \right. \; ,
\end{align*}
where $\vec{n}_1 = (1,0)$ and $\vec{n}_2 = (0,1)$ are the basis
vectors in the space $N$.
Note that some one-dimensional cones $v_i$ are elements of some
two-dimensional cones $\sigma_i$. For example,
$v_1$ and $v_3$ are the elements of $\sigma_1$.
The fan $\Delta$ consists of zero-, one- and two-dimensional cones. 
It is described as Figure \ref{fan-P2}:
\begin{figure}[h]
\begin{center}
\psfrag{(1,0)}{$(1,0)$}
\psfrag{(0,1)}{$(0,1)$}
\psfrag{(-1,-1)}{$(-1,-1)$}
\psfrag{s1}{\large $\sigma_1$}
\psfrag{s2}{\large $\sigma_2$}
\psfrag{s3}{\large $\sigma_3$}
\psfrag{v1}{\large $v_1$}
\psfrag{v2}{\large $v_2$}
\psfrag{v3}{\large $v_3$}
\includegraphics[width=4.5cm]{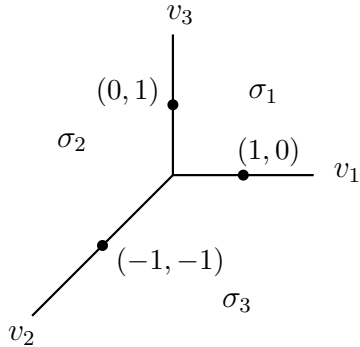}
\caption{\sl Toric fan of the projective space $\P{2}$ in $N_{\R}$.}
\label{fan-P2}
\end{center}
\end{figure}

\noi
{}From the definition of dual cones in (\ref{dual-cones}),
we describe two-dimensional dual cones $\check{\sigma}_i$ as follows:
\begin{align*}
\check{\sigma}_1 \cap M_{\R} 
\ &= \ 
{\R}_{\geq 0} \, {\mbf m}_{11} + {\R}_{\geq 0} \, {\mbf m}_{12} \; , \ls
{\mbf m}_{11} \ = \ \vec{m}_1 \;, \ \ \ {\mbf m}_{12} \ = \ \vec{m}_2 \; , \\
\check{\sigma}_2 \cap M_{\R} 
\ &= \ 
{\R}_{\geq 0} \, {\mbf m}_{21} + {\R}_{\geq 0} \, {\mbf m}_{22} \; , \ls
{\mbf m}_{21} \ = \ - \vec{m}_1 \;, \ \ \ {\mbf m}_{22} \ = \ -
\vec{m}_1 + \vec{m}_2 \; , \\ 
\check{\sigma}_3 \cap M_{\R} 
\ &= \ 
{\R}_{\geq 0} \, {\mbf m}_{31} + {\R}_{\geq 0} \, {\mbf m}_{32} \; , \ls
{\mbf m}_{31} \ = \ \vec{m}_1 - \vec{m}_2 \;, \ \ \ {\mbf m}_{32} \ = \
- \vec{m}_2 \; ,
\end{align*}
where ${\mbf m}_{ij}$ are elements of dual cones $\check{\sigma}_i$.
Note that $\vec{m}_1 = (1,0)$ and $\vec{m}_2 = (0,1)$ are basis
vectors in dual lattice $M$.
Since there are some relations among these elements represented in
(\ref{transit}),
we obtain transition rules among local coordinates $u_{ij}$ in local
patches $U_{\sigma_i}$: 
\begin{align*}
\left\{
\begin{array}{l}
{\mbf m}_{21} + {\mbf m}_{11} \ = \ 0 \\
{\mbf m}_{22} + {\mbf m}_{11} - {\mbf m}_{12} \ = \ 0 
\end{array} \right.
\ &\to \ \ \left\{
\begin{array}{l}
u_{21} \ = \ (u_{11})^{-1} \\
u_{22} \ = \ (u_{11})^{-1} u_{12} 
\end{array} \right. \ls \mbox{in} \ \ U_{\sigma_1} \cap U_{\sigma_2} \;, \\
\left\{
\begin{array}{l}
{\mbf m}_{31} - {\mbf m}_{11} + {\mbf m}_{12} \ = \ 0 \\
{\mbf m}_{32} + {\mbf m}_{12} \ = \ 0 
\end{array} \right.
\ &\to \ \ \left\{
\begin{array}{l}
u_{31} \ = \ u_{11} (u_{12})^{-1} \\
u_{32} \ = \ (u_{12})^{-1} 
\end{array} \right. \ls \mbox{in} \ \ U_{\sigma_1} \cap U_{\sigma_3} \;,
\end{align*}
where $(u_{i1}, u_{i2})$ are local coordinates in $U_{\sigma_i}$. We
summarize these transition rules in Figure \ref{transition-P2}:

\vspace{5mm}

\begin{figure}[h]
\psfrag{1}{$(x,y) \in U_{\sigma_1}$}
\psfrag{2}{$(x^{-1}, x^{-1} y) \in U_{\sigma_2}$}
\psfrag{3}{$(xy^{-1}, y^{-1}) \in U_{\sigma_3}$}
\LS \LS \includegraphics[width=9cm]{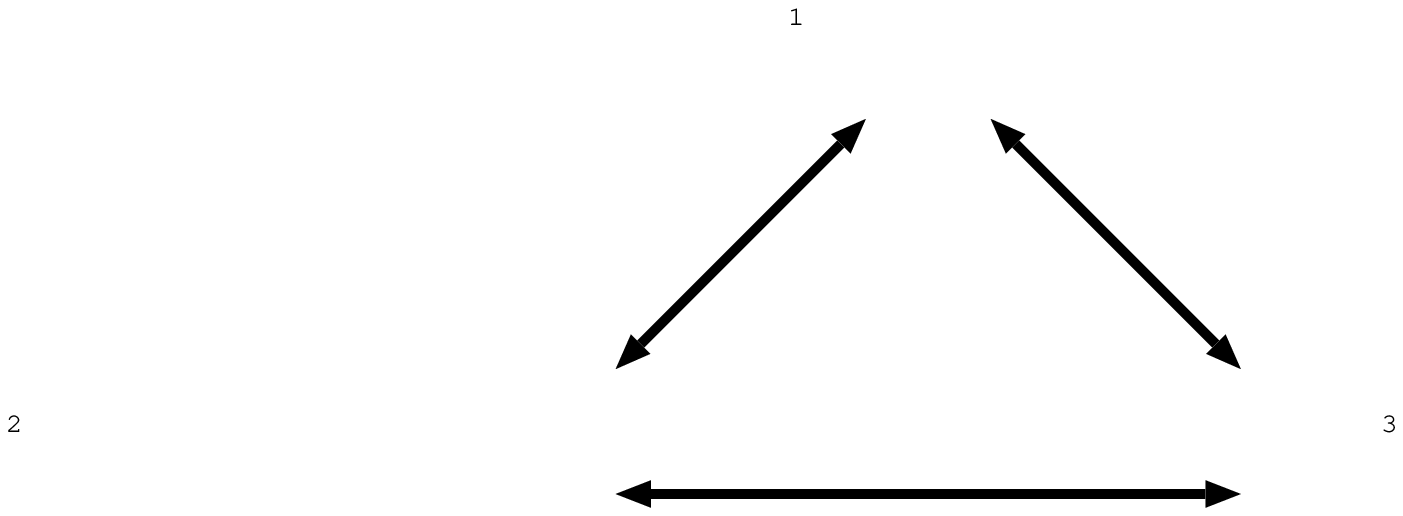}
\begin{center}
\caption{\sl Transition rules among local patches of $\P{2}$.}
\label{transition-P2}
\end{center}
\end{figure}

Let us define the homogeneous coordinates $\{ z_i \}$ 
of $\P{2}$ and consider the relations with local coordinates $\{u_{ij}
\}$ in the local patch $U_{\sigma_i}$.
In terms of (\ref{torus}) we find the following relation among primitive
generators of one-dimensional cones:
\begin{align*}
n(v_1) + n(v_2) + n (v_3) \ &= \ 0 \; \ls
\mbox{where} \ \ \
\left\{
\begin{array}{l}
n (v_1) \ = \ \vec{n}_1 \\
n (v_2) \ = \ - \vec{n}_1 - \vec{n}_2 \\
n (v_3) \ = \ \vec{n}_2 
\end{array} \right. \; .
\end{align*}
Thus we define the homogeneous coordinates $z_i$ and the integers
$q_i$ of $\P{2}$ as Table \ref{homo-charge-P2}:
\begin{table}[h]
\begin{center}
\begin{tabular}{c|ccc}
& $z_1$ & $z_2$ & $z_3$ \\ \hline 
$q_i$ & 1 & 1 & 1 
\end{tabular}
\caption{\sl Homogeneous coordinates and integers of $\P{2}$.}
\label{homo-charge-P2}
\end{center}
\end{table}

\noi
Note that we regard homogeneous coordinates $z_i$ and integers $q_i$
as chiral superfields $\Phi^i$ and their $U(1)$ charges $Q_i$
in the GLSM (see Table \ref{charge-P2}).

We describe the relationships between homogeneous coordinates $z_i$ and
local coordinates $u_{ij}$ in $U_{\sigma_i}$ in terms of (\ref{local-homo}):
\begin{align*}
(u_{11}, u_{12}) \ &= \ \Big( \frac{z_1}{z_2} , \frac{z_3}{z_2} \Big) 
\ls \mbox{in} \ \ U_{\sigma_1} \; , \\
(u_{21}, u_{22}) \ &= \ \Big( \frac{z_2}{z_1} , \frac{z_3}{z_1} \Big) 
\ls \mbox{in} \ \ U_{\sigma_2} \; , \\
(u_{31}, u_{32}) \ &= \ \Big( \frac{z_1}{z_3}, \frac{z_2}{z_3} \Big) 
\ls \mbox{in} \ \ U_{\sigma_3} \; .
\end{align*}


\subsection{Hirzebruch surfaces} \label{fan-def-F}

Let us consider a toric fan of a Hirzebruch surface $\F_k$.
It consists of a zero-dimensional cone, one-dimensional cones $v_i$
and two-dimensional cones $\sigma_i$ which are defined as
\begin{align*}
\left\{
\begin{array}{l}
v_1 \cap N_{\R} \ = \ {\R}_{\geq 0} \, \vec{n}_1 \\
v_2 \cap N_{\R} \ = \ {\R}_{\geq 0} ( - \vec{n}_1 - \vec{n}_2)  \\
v_3 \cap N_{\R} \ = \ {\R}_{\geq 0} \, \vec{n}_2 \\
v_4 \cap N_{\R} \ = \ {\R}_{\geq 0} \, ( - \vec{n}_2)  
\end{array} \right. \; ,
\ls & \ls
\left\{
\begin{array}{l}
\sigma_1 \cap N_{\R} 
\ = \ {\R}_{\geq 0} \, \vec{n}_1 + {\R}_{\geq 0} \, \vec{n}_2 \\
\sigma_2 \cap N_{\R} 
\ = \ {\R}_{\geq 0} \, \vec{n}_2 + {\R}_{\geq 0} (- \vec{n}_1- \vec{n}_2)  \\
\sigma_3 \cap N_{\R} 
\ = \ {\R}_{\geq 0} (- \vec{n}_2)  + {\R}_{\geq 0} (- \vec{n}_1
- \vec{n}_2)  \\
\sigma_4 \cap N_{\R} 
\ = \ {\R}_{\geq 0} \, \vec{n}_1 + {\R}_{\geq 0} ( - \vec{n}_2 )  
\end{array} \right. \; .
\end{align*}
The fan is a collection of these cones. 
We describe the fan in Figure \ref{fan-Hir}:
\begin{figure}
\begin{center}
\psfrag{(1,0)}{$(1,0)$}
\psfrag{(0,1)}{$(0,1)$}
\psfrag{(0,-1)}{$(0,-1)$}
\psfrag{(-1,-a)}{$(-1,-k)$}
\psfrag{s1}{\large $\sigma_1$}
\psfrag{s2}{\large $\sigma_2$}
\psfrag{s3}{\large $\sigma_3$}
\psfrag{s4}{\large $\sigma_4$}
\psfrag{v1}{\large $v_1$}
\psfrag{v2}{\large $v_2$}
\psfrag{v3}{\large $v_3$}
\psfrag{v4}{\large $v_4$}
\includegraphics[width=5.3cm]{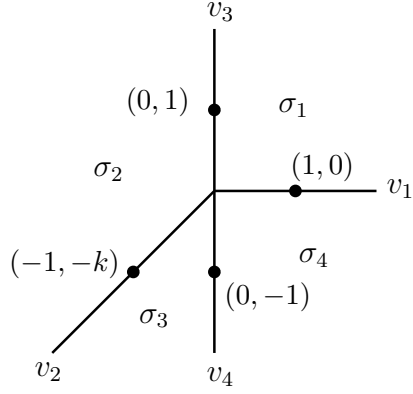}
\caption{\sl Toric fan of the Hirzebruch surface $\F_k$ in $N_{\R}$.}
\label{fan-Hir}
\end{center}
\end{figure}
As we discussed in (\ref{dual-cones}),
two-dimensional dual cones $\check{\sigma}_i$ are defined in terms of
basis vectors $\vec{m}_i$ of dual lattice $M$:
\begin{align*}
\check{\sigma}_1 \cap M_{\R} 
\ &= \ 
{\R}_{\geq 0} \, {\mbf m}_{11} + {\R}_{\geq 0} \, {\mbf m}_{12} \; , \ls
{\mbf m}_{11} \ = \ \vec{m}_1 \;, \ \ \ 
{\mbf m}_{12} \ = \ \vec{m}_2 \; , \\
\check{\sigma}_2 \cap M_{\R} 
\ &= \ 
{\R}_{\geq 0} \, {\mbf m}_{21} + {\R}_{\geq 0} \, {\mbf m}_{22} \; , \ls
{\mbf m}_{21} \ = \ - \vec{m}_1 \;, \ \ \ 
{\mbf m}_{22} \ = \ - k \vec{m}_1 + \vec{m}_2 \; , \\
\check{\sigma}_3 \cap M_{\R} 
\ &= \ 
{\R}_{\geq 0} \, {\mbf m}_{31} + {\R}_{\geq 0} \, {\mbf m}_{32} \; , \ls
{\mbf m}_{31} \ = \ - \vec{m}_1 \;, \ \ \ 
{\mbf m}_{32} \ = \ k \vec{m}_1 - \vec{m}_2 
\; , \\
\check{\sigma}_4 \cap M_{\R} 
\ &= \ 
{\R}_{\geq 0} \, {\mbf m}_{41} + {\R}_{\geq 0} \, {\mbf m}_{42} \; , \ls
{\mbf m}_{41} \ = \ \vec{m}_1 \;, \ \ \ 
{\mbf m}_{42} \ = \ - \vec{m}_2 \; .
\end{align*}
There are some relations among the elements ${\mbf m}_{ij}$.
Thus we construct transition rules from them as
\begin{align*}
&\left\{
\begin{array}{l}
{\mbf m}_{21} + {\mbf m}_{11} \ = \ 0 \\
{\mbf m}_{22} + k {\mbf m}_{11} - {\mbf m}_{12} \ = \ 0 
\end{array} \right.
&\to \ls
&\left\{
\begin{array}{l}
u_{21} \ = \ (u_{11})^{-1} \\
u_{22} \ = \ (u_{11})^{-k} \, u_{12} 
\end{array} \right.
& \mbox{in} \ \ U_{\sigma_1} \cap U_{\sigma_2} \; , \\
&\left\{
\begin{array}{l}
{\mbf m}_{31} + {\mbf m}_{11} \ = \ 0 \\
{\mbf m}_{32} - k {\mbf m}_{11} + {\mbf m}_{12} \ = \ 0 
\end{array} \right.
&\to \ls
&\left\{
\begin{array}{l}
u_{31} \ = \ (u_{11})^{-1} \\
u_{32} \ = \ (u_{11})^k (u_{12})^{-1}  
\end{array} \right.
& \mbox{in} \ \ U_{\sigma_1} \cap U_{\sigma_3} \; , \\
&\left\{
\begin{array}{l}
{\mbf m}_{41} - {\mbf m}_{11} \ = \ 0 \\
{\mbf m}_{42} + {\mbf m}_{12} \ = \ 0
\end{array} \right.
&\to \ls
&\left\{
\begin{array}{l}
u_{41} \ = \ u_{11} \\
u_{42} \ = \ (u_{12})^{-1}
\end{array} \right.
& \mbox{in} \ \ U_{\sigma_1} \cap U_{\sigma_4} \; .
\end{align*}
We denote these transition rules in Figure \ref{transition-F}:

\vspace{5mm}

\begin{figure}[h]
\psfrag{1}{$(x,y) \in U_{\sigma_1}$}
\psfrag{2}{$(x^{-1}, x^{-k} y) \in U_{\sigma_2}$}
\psfrag{3}{$(x^{-1}, x^{k} y^{-1}) \in U_{\sigma_3}$}
\psfrag{4}{$(x, y^{-1}) \in U_{\sigma_4}$}
\LS \LS \ \ 
\includegraphics[width=10cm]{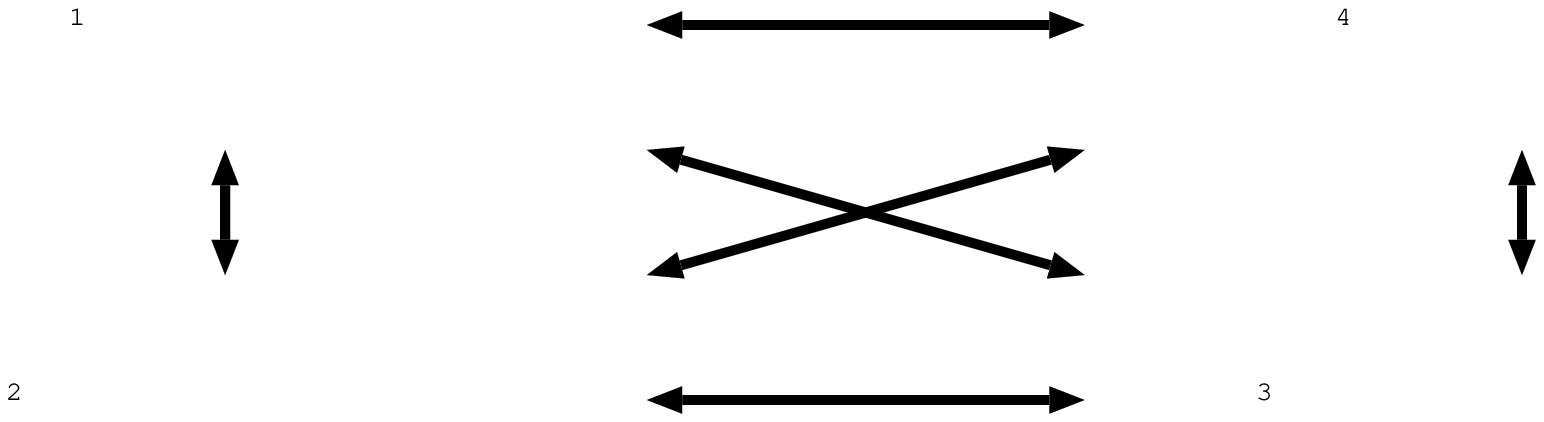}
\begin{center}
\caption{\sl Transition rules among local patches of $\F_k$.}
\label{transition-F}
\end{center}
\end{figure}

The relationships among the primitive generators of one-dimensional
cones are described as
\begin{align*}
n(v_1) + n(v_2) + n (v_3) \ &= \ 0 \; , \ls 
n(v_3) + n(v_4) \ = \ 0 \; ,
\end{align*}
where 
\begin{align*}
n(v_1) \ &= \ \vec{n}_1 \; , \ls
n(v_2) \ = \ - \vec{n}_1 - \vec{n}_2 \; , \ls
n(v_3) \ = \ \vec{n}_2 \; , \ls
n(v_4) \ = \ -\vec{n}_2 \; .
\end{align*}
Due to these relations we write down homogeneous coordinates and
integers as Table \ref{homo-charge-F}:
\begin{table}[h]
\begin{center}
\begin{tabular}{c|cccc}
& $z_1$ & $z_2$ & $z_3$ & $z_4$ \\ \hline
$q_i^1$ & 1 & 1 & $k$ & 0 \\
$q_i^2$ & 0 & 0 & 1 & 1 
\end{tabular}
\caption{\sl Homogeneous coordinates $z_i$ and integers $q_i^a$ of
  $\F_k$.}
\label{homo-charge-F}
\end{center}
\end{table}

\vspace{-7mm}

\noi
Note that we regard homogeneous coordinates $z_i$ and integers $q_i^a$
as chiral superfields $\Phi^i$ and their $U(1) \times U(1)$ charges $Q_i^a$
in the GLSM (see Table \ref{charge-F}).

We denote the relationships between homogeneous coordinates and local
coordinates:
\begin{align*}
(u_{11}, u_{12}) \ &= \ \Big( \frac{z_1}{z_2} , \frac{z_3}{(z_2)^k z_4}
  \Big) \ls \mbox{in} \ \ U_{\sigma_1} \; , \ls
(u_{21}, u_{22}) \ = \ \Big( \frac{z_2}{z_1} , \frac{z_3}{(z_1)^k
  z_4} \Big) \ls \mbox{in} \ \ U_{\sigma_2} \; , \\
(u_{31}, u_{32}) \ &= \ \Big( \frac{z_2}{z_1} , \frac{(z_1)^k
  z_4}{z_3} \Big) \ls \mbox{in} \ \ U_{\sigma_3} \; , \ls
(u_{41}, u_{42}) \ = \ \Big( \frac{z_1}{z_2} , \frac{(z_2)^k
  z_3}{z_4} \Big) \ls \mbox{in} \ \ U_{\sigma_4} \; .
\end{align*}


\subsection{del Pezzo surfaces} \label{fan-def-B2}

Here we consider a toric data of del Pezzo surface $\B_2$. 
We define one- and two-dimensional cones as follows:
\begin{align*}
\left\{
\begin{array}{l}
v_1 \cap N_{\R} \ = \ {\R}_{\geq 0} \, \vec{n}_1 \\
v_2 \cap N_{\R} \ = \ {\R}_{\geq 0} ( - \vec{n}_1 - \vec{n}_2)  \\
v_3 \cap N_{\R} \ = \ {\R}_{\geq 0} \, \vec{n}_2 \\
v_4 \cap N_{\R} \ = \ {\R}_{\geq 0} \, ( - \vec{n}_2)  \\
v_5 \cap N_{\R} \ = \ {\R}_{\geq 0} \, ( - \vec{n}_1)  
\end{array} \right. \; ,
\ls & \ls
\left\{
\begin{array}{l}
\sigma_1 \cap N_{\R} 
\ = \ {\R}_{\geq 0} \, \vec{n}_1 + {\R}_{\geq 0} \, \vec{n}_2 \\
\sigma_2 \cap N_{\R} 
\ = \ {\R}_{\geq 0} \, \vec{n}_2 + {\R}_{\geq 0} (- \vec{n}_1)  \\
\sigma_3 \cap N_{\R} 
\ = \ {\R}_{\geq 0} (- \vec{n}_1)  + {\R}_{\geq 0} (- \vec{n}_1
- \vec{n}_2)  \\
\sigma_4 \cap N_{\R} 
\ = \ {\R}_{\geq 0} (- \vec{n}_1 - \vec{n}_2) 
+ {\R}_{\geq 0} ( - \vec{n}_2 )  \\
\sigma_5 \cap N_{\R} 
\ = \ {\R}_{\geq 0} (- \vec{n}_2) + {\R}_{\geq 0} \, \vec{n}_1
\end{array} \right. \; .
\end{align*}
The toric fan $\Delta$ consists of these cones.
We describe $\Delta$ in Figure \ref{fan-B2}:
\begin{figure}[h]
\begin{center}
\psfrag{(1,0)}{$(1,0)$}
\psfrag{(0,1)}{$(0,1)$}
\psfrag{(-1,0)}{$(-1,0)$}
\psfrag{(0,-1)}{$(0,-1)$}
\psfrag{(-1,-1)}{$(-1,-1)$}
\psfrag{s1}{\large $\sigma_1$}
\psfrag{s2}{\large $\sigma_2$}
\psfrag{s3}{\large $\sigma_3$}
\psfrag{s4}{\large $\sigma_4$}
\psfrag{s5}{\large $\sigma_5$}
\psfrag{v1}{\large $v_1$}
\psfrag{v2}{\large $v_2$}
\psfrag{v3}{\large $v_3$}
\psfrag{v4}{\large $v_4$}
\psfrag{v5}{\large $v_5$}
\includegraphics[width=6cm]{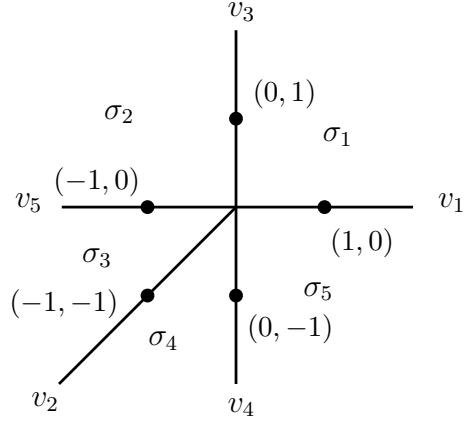}
\caption{\sl Toric fan of the del Pezzo surface $\B_2$ in $N_{\R}$.}
\label{fan-B2}
\end{center}
\end{figure}

The dual cones $\check{\sigma}_i$ are defined in terms of basis
vectors $\vec{m}_i$:
\begin{align*}
\check{\sigma}_1 \cap M_{\R} 
\ &= \ 
{\R}_{\geq 0} \, {\mbf m}_{11} + {\R}_{\geq 0} \, {\mbf m}_{12} \; , \ls
{\mbf m}_{11} \ = \ \vec{m}_1 \;, \ \ \ 
{\mbf m}_{12} \ = \ \vec{m}_2 \; , \\
\check{\sigma}_2 \cap M_{\R} 
\ &= \ 
{\R}_{\geq 0} \, {\mbf m}_{21} + {\R}_{\geq 0} \, {\mbf m}_{22} \; , \ls
{\mbf m}_{21} \ = \ - \vec{m}_1 \;, \ \ \ 
{\mbf m}_{22} \ = \ \vec{m}_2 \; , \\
\check{\sigma}_3 \cap M_{\R} 
\ &= \ 
{\R}_{\geq 0} \, {\mbf m}_{31} + {\R}_{\geq 0} \, {\mbf m}_{32} \; , \ls
{\mbf m}_{31} \ = \ - \vec{m}_1 + \vec{m}_2 \;, \ \ \ 
{\mbf m}_{32} \ = \ - \vec{m}_2
\; , \\
\check{\sigma}_4 \cap M_{\R} 
\ &= \
{\R}_{\geq 0} \, {\mbf m}_{41} + {\R}_{\geq 0} \, {\mbf m}_{42} \; , \ls
{\mbf m}_{41} \ = \ \vec{m}_1 - \vec{m}_2 \;, \ \ \ 
{\mbf m}_{42} \ = \ - \vec{m}_1 
\; , \\
\check{\sigma}_5 \cap M_{\R} 
\ &= \
{\R}_{\geq 0} \, {\mbf m}_{51} + {\R}_{\geq 0} \, {\mbf m}_{52} \; , \ls
{\mbf m}_{51} \ = \ \vec{m}_1 \;, \ \ \ 
{\mbf m}_{52} \ = \ - \vec{m}_2
\; .
\end{align*}
In terms of these data we determine the transition rules of local
coordinates among local patches:
\begin{align*}
&\left\{
\begin{array}{l}
{\mbf m}_{21} + {\mbf m}_{11} \ = \ 0 \\
{\mbf m}_{22} - {\mbf m}_{12} \ = \ 0 
\end{array} \right.
&\to \ls
&\left\{
\begin{array}{l}
u_{21} \ = \ (u_{11})^{-1} \\
u_{22} \ = \ u_{12} 
\end{array} \right.
& \mbox{in} \ \ U_{\sigma_1} \cap U_{\sigma_2} \; , \\
&\left\{
\begin{array}{l}
{\mbf m}_{31} + {\mbf m}_{11} - {\mbf m}_{12} \ = \ 0 \\
{\mbf m}_{32} + {\mbf m}_{12} \ = \ 0 
\end{array} \right.
&\to \ls
&\left\{
\begin{array}{l}
u_{31} \ = \ (u_{11})^{-1} \, u_{12} \\
u_{32} \ = \ (u_{12})^{-1}  
\end{array} \right.
& \mbox{in} \ \ U_{\sigma_1} \cap U_{\sigma_3} \; , \\
&\left\{
\begin{array}{l}
{\mbf m}_{41} - {\mbf m}_{11} + {\mbf m}_{12} \ = \ 0 \\
{\mbf m}_{42} + {\mbf m}_{11} \ = \ 0
\end{array} \right.
&\to \ls
&\left\{
\begin{array}{l}
u_{41} \ = \ u_{11} (u_{12})^{-1} \\
u_{42} \ = \ (u_{11})^{-1}
\end{array} \right.
& \mbox{in} \ \ U_{\sigma_1} \cap U_{\sigma_4} \; , \\
&\left\{
\begin{array}{l}
{\mbf m}_{51} - {\mbf m}_{11} \ = \ 0 \\
{\mbf m}_{52} + {\mbf m}_{12} \ = \ 0
\end{array} \right.
&\to \ls 
&\left\{
\begin{array}{l}
u_{51} \ = \ u_{11} \\
u_{52} \ = \ (u_{12})^{-1}
\end{array} \right.
& \mbox{in} \ \ U_{\sigma_1} \cap U_{\sigma_5} \; .
\end{align*}
where $(u_{i1}, u_{i2})$ are local coordinates in $U_{\sigma_i}$.
We summarize these transition rules in Figure \ref{transition-B}:

\vspace{5mm}

\begin{figure}[h]
\psfrag{1}{$(x,y) \in U_{\sigma_1}$}
\psfrag{2}{$(x^{-1}, y) \in U_{\sigma_2}$}
\psfrag{3}{$(x^{-1} y, y^{-1}) \in U_{\sigma_3}$}
\psfrag{4}{$(x y^{-1}, x^{-1}) \in U_{\sigma_4}$}
\psfrag{5}{$(x, y^{-1}) \in U_{\sigma_5}$}
\LS \ls \includegraphics[width=11cm]{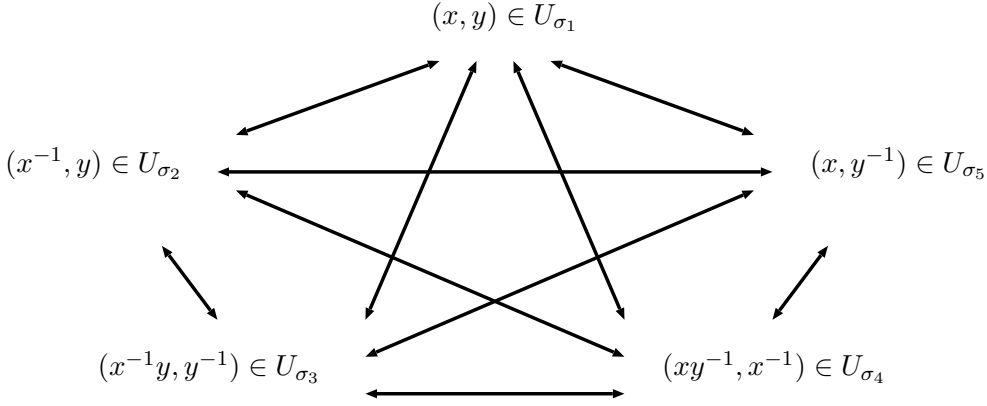}
\begin{center}
\caption{\sl Transition rules among local patches of $\B_2$.}
\label{transition-B}
\end{center}
\end{figure}

There are some relationships among the primitive generators:
\begin{align*}
\left\{
\begin{array}{l}
n(v_1) + n(v_2) + n(v_3) \ = \ 0 \\
n(v_3) + n(v_4) \ = \ 0 \\
n(v_1) + n(v_5) \ = \ 0 
\end{array} \right.
\ls \mbox{where} \ls \left\{
\begin{array}{l}
n (v_1) \ = \ \vec{n}_1 \\
n (v_2) \ = \ - \vec{n}_1 - \vec{n}_2 \\
n (v_3) \ = \ \vec{n}_2 \\
n (v_4) \ = \ - \vec{n}_2 \\
n (v_5) \ = \ - \vec{n}_1 
\end{array} \right. \; .
\end{align*}
Due to these relations 
we define the homogeneous coordinates and integers as Table
\ref{homo-charge-B}:
\begin{table}[h]
\begin{center}
\begin{tabular}{c|ccccc}
& $z_1$ & $z_2$ & $z_3$ & $z_4$ & $z_5$ \\ \hline
$q_i^1$ & 1 & 1 & 1 & 0 & 0 \\
$q_i^2$ & 0 & 0 & 1 & 1 & 0 \\
$q_i^3$ & 1 & 0 & 0 & 0 & 1
\end{tabular}
\caption{\sl Homogeneous coordinates $z_i$ and integers $q_i^a$ of $\B_2$.}
\label{homo-charge-B}
\end{center}
\end{table}

\vspace{-7mm}

\noi
Note that we regard homogeneous coordinates $z_i$ and integers $q_i^a$
as chiral superfields $\Phi^i$ and their $U(1)^3$ charges $Q_i^a$
in the GLSM (see Table \ref{charge-B}).

We discuss the relationships between local coordinates and homogeneous
coordinates in terms of (\ref{local-homo}) as follows:
\begin{align*}
(u_{11}, u_{12}) \ &= \ \Big( \frac{z_1}{z_2 z_5} , \frac{z_3}{z_2 z_4}\Big) 
\ls \mbox{in} \ \ U_{\sigma_1} \;  , \ls 
(u_{21}, u_{22}) \ = \ \Big( \frac{z_2 z_5}{z_1} , \frac{z_3}{z_2 z_4} \Big)
 \ls \mbox{in} \ \ U_{\sigma_2} \; , \\
(u_{31}, u_{32}) \ &= \ \Big( \frac{z_3 z_5}{z_1 z_4} , 
\frac{z_2 z_4}{z_3} \Big) 
\ls \mbox{in} \ \ U_{\sigma_3} \;  , \ls
(u_{41}, u_{42}) \ = \ \Big( \frac{z_1 z_4}{z_3 z_5} , \frac{z_2
  z_5}{z_1} \Big) 
\ls \mbox{in} \ \ U_{\sigma_4} \;  , \\
(u_{51}, u_{52}) \ &= \ \Big( \frac{z_1}{z_2 z_5} , \frac{z_2 z_4}{z_3} \Big) 
\ls \mbox{in} \ \ U_{\sigma_5} \; .
\end{align*}

\end{appendix}



}
\end{document}